\documentclass[epj]{svjour}
\usepackage{graphics}
\usepackage{notoccite}
\begin{document}
\title{A statistical model to calculate inclusive hadronic cross sections }
\author{Gábor Balassa\inst{1} \and György Wolf \inst{1} 
}                     
%
%
\institute{\inst{1}Institute for Particle and Nuclear Physics, Wigner Research Centre for Physics, H-1525 Budapest, Hungary}
\date{Received: date / Revised version: date}
%
\abstract{
Hadronic cross sections are important ingredients in many of the ongoing research methods in high energy nuclear physics, and it is always important to measure and/or calculate the probabilities of different types of reactions. In heavy-ion transport simulations at a few GeV energies, these hadronic cross sections are essential and so far mostly the exclusive processes are used, however, if one interested in total production rates the inclusive cross sections are also necessary to know. In this paper, we introduce a statistical-based method, which is able to give good estimates to exclusive and inclusive cross sections as well in the energy range of a few GeV. The method and its estimates for not well-known cross sections, will be used in a Boltzmann-Uehling-Uhlenbeck (BUU) type off-shell transport code to explain charmonium and bottomonium mass shifts in heavy-ion collisions.
\PACS{
      {PACS-key}{discribing text of that key}   \and
      {PACS-key}{discribing text of that key}
     } 
} 
\maketitle
\section{Introduction}
\label{intro}
Hadronic cross sections at a few GeV are usually used as input parameters in heavy-ion transport codes, where up to $10-20$ GeV the main degrees of freedom are the known baryons, mesons and the well-established resonances  \cite{BATKO1991331,EHEHALT199331,Maruyama:1993jb,PhysRevC.97.034625}. 
In these simulations mostly $2 \rightarrow 2$ exclusive reactions are used as input, however there were some attempts to include $N \rightarrow M$ reactions as well \cite{BATKO1992786,CASSING2002618}. It is not straightforward how to include these types of collisions into the codes, so these are mostly omitted in the simulations. At higher energies, partonic degrees of freedoms are also becoming important \cite{Bratkovskaya:2011wp}. Recent calculations show that the charmonium states e.g. $J/\Psi , \Psi(3686), \Psi(3770)$ will acquire mass shifts during their evolution in dense matter, which could indicate a non-zero value for the gluon condensate at specific nuclear densities \cite{M1,M2,M3}. This information, if measured could be really important to non-perturbative quantum chromodynamics. For this purpose the inclusive charmonium production cross sections should be known, however, they are not well (or not at all) measured in the energy range we are interested in (2-10 GeV), so it has to be estimated from some theory. The method presented in this paper is able to predict inclusive cross sections, which will be ultimately used to estimate charmonium production cross sections. The paper is organized as follows. In sections one and two a basic formulation of the method and its capabilities to estimate exclusive cross sections are described. This can be found more rigorously in \cite{Balassa:2017pgf}, where the main idea and some examples were also given. After the basic formulation, section three dedicated to the normalization and model error estimation. The fourth section describes a method, which is used to calculate inelastic cross sections without summing over all the possible $2 \rightarrow N$ reactions and using only 1-, or 2 fireball event ratios. The model error is also addressed in this section. The fifth section divided into two subsections. In the first subsection six examples are given for the inclusive cross sections compared to their measured values, while the second subsection corresponds to charmonium production and the estimation of the charm quark creational probability. Finally the last section briefly summarizes the paper.
\section{Basic formulation and normalization}
\label{sec:1}
The method is based on the so-called Statistical Bootstrap \cite{H1,H2,H3} approach, which idea was widely used to estimate particle multiplicities at high energies  \cite{Mult1,Mult2,Mult3}. In our model, we assume that at the initial state of a collision a strongly interacting, compound system is formed (a so-called fireball), which will ultimately decay into hadrons giving some specific final state of the collision. In the reaction probability, the initial fireball formation stage and the hadronization stage can be separated as it is shown in Eq.\ref{eq:1}, where we assume that the first stage is described by the total or inelastic cross sections\footnote{In our model it is the inelastic cross section is used, rather than the total due to the exclusion of the elastic channel.} of the reaction and the second stage is described by some mixed statistical and dynamical factors.
\begin{eqnarray}
\label{eq:1}
\nonumber
&{\displaystyle \sigma^{n \rightarrow k}(E)=\Bigg( \int \prod_{i=1}^{n} d^3p_i R(E,p_1,...,p_n) \Bigg) } \\
&{\displaystyle \times \Bigg( \int \prod_{i=1}^{k} d^3q_i w(E,q_1,...,q_k) \Bigg)}
\end{eqnarray}
where $E$ is the CM energy of the collision ($E=\sqrt{s}$), $\sigma^{n \rightarrow k}$ is the generalized n-body cross section, $p_1,...p_n$ are the momenta of the incoming particles, $q_1,...q_k$ are the momenta of the outgoing particles, $R(E,...)$ is the function describing the initial state dynamics, and $w(E,...)$ describes the probability of a specific k-body final state. At the first stage of the hadronization process the fireball with invariant mass $M=\sqrt{s}$ directly giving hadrons or decay into two subsequent fireballs with smaller masses $m_1,m_2$ where $m_1+m_2=M$ should be fulfilled. The resulting fireball will hadronize or decay into further fireballs. At the end of the chain, all of the fireballs have to hadronize leaving only hadrons at the end. The probability of the one-, two, three-, etc... fireball scheme can be calculated with an appropriate model described in our previous paper on this topic \cite{Balassa:2017pgf}. The possible number of hadrons coming from one fireball is two or three\footnote{There is also a possibility that one fireball could decay into more than three particles, however due to the small probabilities, they are neglected in our model.} with the probabilities of $P_2^d=0.69$, $P_3^d=0.24$ which is described in the original formulation of the statistical bootstrap by Frautschi \cite{PhysRevD.3.2821}.

An extra ingredient to the full probability (not mentioned in the original formulation, because the normalization was cancelled in the calculated ratios), which is needed to make a proper normalization is to separate stable and non-stable/resonant particles and use a weighting factor to each of the channels corresponding to the full width of the particles. In the model, the stable particles are the ones, which could not decay into other hadrons via the strong interaction. Examples are the $N^*$ nucleon resonances, which also could decay strongly to protons/neutrons, so they are not stable. These unstable particles are weighted by a relativistic Breit-Wigner factor\cite{BR}:
\begin{equation}
\label{eq:2}
F_i^{BR}(E_i,m_i) = \frac{1}{\pi}\frac{E_i^2 \Gamma_i}{(E_i^2-m_i^2)^2+E_i^2 \Gamma_i^2}
\end{equation}
where $m_i$ is the $i$'th particle mass, $\Gamma_i$ is its total width, $E_i$ is the invariant mass of the resonance, and the energy integral of $F_i^{BR}(E_i,m_i)$ is normalized to unity in $E\in[-\infty,\infty]$.

The appearance of the Breit-Wigner factor can be explained by the propagator of a particle with finite but non negligible width, when calculating Feynman diagrams e.g. as it was described in \cite{DIM}. To calculate it's contribution to the full probbility it has to be integrated out in the physical region, defined by the masses of the possible decay products of each resonance. 

Putting together everything the full probability of a one-fireball scheme, which hadronizes into two or three particles can be calculated as it is shown in Eq.\ref{eq:3}:
\begin{eqnarray}
\label{eq:3}
\nonumber
&{\displaystyle W_{1,i}^{n_i}(E)=P_1^{fb}(E) \frac{ C_{Q_i}(E)P_{n_i}^{H,i}(E)}{\sum_j C_{Q_j}(E)P_{n_j}^{H,j}(E)}  = } \\
&{\displaystyle = P_1^{fb}(E) \frac{T_i(E)}{\sum_j T_j(E)}},
\end{eqnarray}
where $P_1^{fb}$ is the probability of the one fireball decay scheme, $C_{Q_i}$ is the quark combinatorial probability for final state $i$, which is proportional to the number of quark-antiquark pairs created at energy $E$, and to the $u,d,s,c,...$ quark creational probabilities fitted from measured data. Furthermore, $n_i$ stands for the number of hadrons coming from the $i$'th fireball (2 or 3) and $P_{n_i}^{H,1}$ is the hadronization probability factor given by Eq.\ref{eq:4} and Eq.\ref{eq:5} for two- and three-body states respectively:
\begin{equation}
\label{eq:4}
P_{2}^{H,i}(E) = \prod_{l=1}^2 (2s_l+1)P_2^d \frac{\Phi_2(E,m_1,m_2)}{\rho(E)(2\pi)^3 N_I!}
\end{equation}
\begin{equation}
\label{eq:5}
P_{3}^{H,i}(E) = \prod_{l=1}^3 (2s_l+1)P_3^d \frac{\Phi_3(E,m_1,m_2,m_3)}{\rho(E)(2\pi)^6 N_I!},
\end{equation}
where $P_2^d=0.69$ and  $P_3^d=0.24$ are the probabilities of a two or three particle final state decays, $s_l$ is the spin of the $l$'th particle, $N_I$ is the number of same particles in the final state, $m_i$ are the masses of particles,  $\Phi_n(E)$ is the $n$-body phase space integral, ans $\rho(E)$ is the density of states (DOS) given by the statistical bootstrap \cite{RHO1,RHO2} :
\begin{equation}
\label{eq:52}
\rho(E) = \frac{a\sqrt{E}}{(E_0+E)^{3.5}}e^{\frac{E}{T_0}},
\end{equation}
where $a$,$E_0$ and $T_0$  are free parameters, however $a$ always factorized out due to the normalization and $E_0=500$ MeV is previously fitted to the experimentally measured DOS \cite{RHO3}. In our previous paper we also set $T_0=160$ MeV using calculated and measured cross section ratios.

All of the factors in Eq.\ref{eq:3} are described in detail in \cite{Balassa:2017pgf}. To simplify the following sections we have introduced the notation:
\begin{equation}
\label{eq:6}
T_i(E) =  P_i^{fb}(E)C_{Q_i}(E)P_{n_i}^{H,i}(E)
\end{equation}

\begin{figure}
\resizebox{0.5\textwidth}{!}{%
  \includegraphics{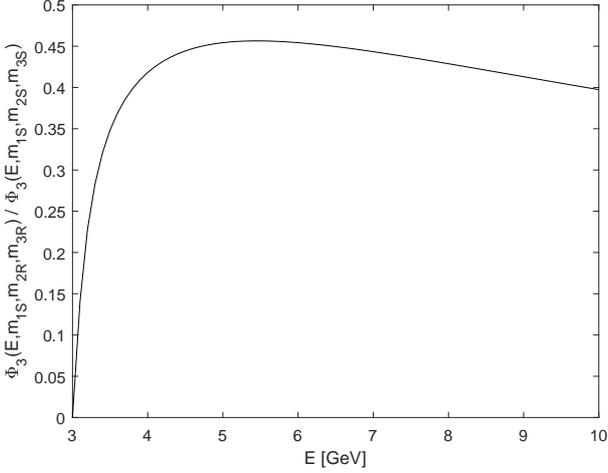}
}
\caption{Ratio of the three-body phase-space integrals for one resonant and two stable vs. three stable particle configurations with $m_1=m_2=m_3=1$ GeV masses, and $\Gamma=0.2$ GeV width for the resonance. The $S$ index in $m_{iS}$ stands for "Stable", while the $R$ index stands for 'Resonant' particles.}
\label{fig:1}       
\end{figure}

The two- and three-body phase-spaces are weighted by the product of the particles Breit-Wigner factors (if it is a resonance) and integrated out to the corresponding energy region as it is seen in Eq.\ref{eq:7}:
\begin{eqnarray}
\label{eq:7}
\nonumber
&{\displaystyle\Phi_k(M,m_1,..,m_k) = V^{k-1} \times} \\
& {\displaystyle\Big( \int \prod_{r \in R} dE_r \Big) \Big( \int \prod_{i=1}^k d^3 \mathbf{q}_i \Big) \prod_{r \in R}^{}F_r^{BR}(E_r,m_r)  \times} \nonumber \\
& {\displaystyle \delta \Big( \sum_{j=1}^k E_j - M \Big) \delta \Big( \sum_{j=1}^k \mathbf{q}_j \Big)}
\end{eqnarray}
where $\mathbf{q}_j$ is the three momenta of the j'th particle, $E_i$ is the energy of the i'th particle and $F_r^{BR}(E_r,m_r) $ is the Breit-Wigner probability factor of the resonance $r$, and $V$ is the interaction volume. If all of the particles are stable then the integrals in the first bracket and the Breit-Wigner factors are not needed. Otherwise the $r$ index goes through all of the resonances. The Breit-Wigner factor will decrease the phase-space according to the resonance width e.g. for a three-body channel with one resonance and two stable particles ($m_1=m_2=m_3=1$ GeV and $\Gamma_1=0.2$ GeV) the corresponding phase-space ratio compared to the all stable particle configuration can be seen in Fig.\ref{fig:1}.
The phase-space with a resonant particle is decreased accordingly to the Breit-Wigner factors inside the phase-space integrals.

The normalization sum in Eq.$\ref{eq:3}$ goes through all the possible one-fireball decays which are allowed by the initial-state quantum numbers. This implies that the normalization sum is different for each reaction e.g. it is not the same for $p\bar{p}$ and $\pi^- p$ reactions due to their different quantum numbers. The extension to more fireballs is not straightforward, due to the different possibilities of how to include quantum number conservation into the model. In our normalization scheme, only the full final state has to respect the conservation laws and the quantum numbers of the different fireballs could be anything, there are no restrictions to it. The normalization procedure for more than one fireball is straightforward to see from the two fireball probability in Eq.\ref{eq:8}:
\begin{equation}
\label{eq:8}
W_{2,ij}^{n_i,n_j}(E)=P_2^{fb}(E) \frac{\frac{1}{N_{i,j}!} \int_{x_{min}}^{x_{max}} dx \frac{T_i(x)}{\sum_a T_a(x)} \frac{T_j(E-x)}{{\sum_a T_a(E-x)}}}{Z_2(E)}
\end{equation}
where the normalization sum $Z_2(E)$ is described by Eq.\ref{eq:9}:
\begin{equation}
\label{eq:9}
Z_2(E) = \sum_{<kl> \in S}\frac{1}{N_{k,l}!} \int_{x_{min}}^{x_{max}} dx \frac{T_k(x)}{{\sum_a T_a(x)}} \frac{T_l(E-x)}{{\sum_a T_a(E-x)}}
\end{equation}
and $<ij>$ means a specific hadronic channel for each fireball e.g. $i=\pi^+ \pi^0$ for the first fireball, and $j=pn$ for the second fireball, so the expression gives us the $X \rightarrow \pi^+\pi^0 pn$ probability. $P_2^{fb}$ is the probability of two fireball formation, $x_{min}=2 m_{\pi^0}$ is the threshold energy, $i,j,k,l$ means a specific final state e.g. $i=p\bar{p}\pi^0$ so $n_i=3$, $N_{i,j}=1$ if the two fireballs give different final states, and $N_{i,j}=2$ if the two fireballs are the same $i=j$. In the normalization sum in the denominator $kl \in S$ means it goes through all the possible final states, whose quantum numbers equal the initial state quantum numbers. The normalization sum for $W_1$ goes through all the possible two-, and three-body final states regardless of the quantum numbers of the initial state. A short example is the $p\bar{p}$ collision, where we have $S=(0,0,0,0)$ (baryon number, charge, strangeness, charmness). For the two fireball normalization sum, we only consider the channels which at the end give back the initial quantum numbers e.g. $Q_1=(1,0,-1,0)$ and $Q_2=(-1,0,1,0)$ is a good choice as $S=Q_1+Q_2$. So separately the fireballs do not have to respect the conservation laws, but the whole system has to and give back the initial quantum numbers.
It is important to note that the N-fireball probability can be built up by the 1-fireball factors and their corresponding integrals as it can be seen in Eq.\ref{eq:10} for the general case with N-fireballs.
\begin{eqnarray}
\label{eq:10}
\nonumber
&{\displaystyle W_{k,i_1..i_k}(E) =P_k^{fb}(E)  \frac{1}{Z_k}\frac{1}{N_{i_1,..i_k}!}\int_{x_{min}}^{x_{max}} \prod_{a=1}^{k} } \Bigg[ dx_a  \times \\ 
& {\displaystyle\times \frac{T_{i_a}(x)}{\sum_j T_j(x)}  \delta \Big( \sum_{a=1}^k x_a-E \Big)  \Bigg]}
\end{eqnarray}
where $Z_k(E)$ is the energy-dependent k-fireball normalization factor given by Eq.\ref{eq:11}.
\begin{eqnarray}
\label{eq:11}
&{\displaystyle Z_k(E)=  \sum_{<l_1..l_k> \in S} \frac{1}{N_{l_1,..l_k}!} \int_{x_{min}}^{x_{max}} \prod_{a=1}^{k} } \Bigg[ dx_a  \times \\ \nonumber
& {\displaystyle\times \frac{T_{l_a}(x)}{\sum_j T_j(x)}  \delta \Big( \sum_{a=1}^k x_a-E \Big)  \Bigg]}
\end{eqnarray}
The complicated normalization factors can be dropped if the ratio of the two process probabilities are taken with strictly the same number of fireballs. If one channel could come from more fireball decays e.g. $2\pi^+2\pi^-$, which could come from one fireball $X\rightarrow \rho^0\rho^0$ and two fireball decay as well $(X_1, X_2)\rightarrow(2\pi^+,2\pi^-)$ the normalization factors do not drop out, however, if we have two channels, which exclusively come from 1 fireball decay e.g. $X\rightarrow \pi^+ \pi^-$ and $X\rightarrow n \bar{n}$, then taking the ratio of the two probabilities the normalization factor is dropped out. If we take the measured cross section for one of the channels and multiply it with the calculated ratio from the model, we get back the unknown cross section of the other channel. This practically only works for low multiplicities, as from one fireball the maximum number of hadrons, we can get is very limited. 

To conclude this section the model is shown to be able to give reasonably good estimates to cross sections with calculating the probability ratios as it is shown in our previous paper. The only free parameters in the model are the quark creational probabilities, the interaction volume $V$, and the hadronization temperature $T_0$. The density of states contains two more previously fitted parameters $a, E_0$, where $a$ can always be factored out by normalization, and $E_0$ is fixed by fitting to the experimentally observed DOS. In our previous paper, we fixed the temperature to $T_0=160$ MeV by calculating cross section ratios. The interaction volume is set to a corresponding interaction radius of $r=0.5$ fm, which describes the calculated inclusive cross sections better at higher energies than a larger value. The quark creational probabilities are hidden in the energy-dependent quark combinatorial probability factor. Their fitted values in the energy range of $E=2-8$ GeV for the three lightest quarks are $P_u=P_d=0.425$ and $P_s=0.15$. These fits were made without including charmed particles, however due to the smallness of the charm creational probabilities, these values are approximately still valid in the energy range, we are interested in. For the heavier charm quarks and the corresponding charmonium particles, the fit $P_c$ is not straightforward as the experimental cross sections are very limited and the cross sections are too small to use the same method which is used to fit $P_u, P_d, P_s$. The solution to this problem is described in Sec.\ref{sec:42}.

There are some limitations to the model, however, which has to be addressed. It is not suited to describe the elastic part of the two-body reactions, so that channel is simply omitted from the calculations and the inelastic ratio is used instead of the total cross section. Another limitation is that the model cannot describe the $A+B \rightarrow C$ reactions, where $A,B,C$ are some particles. This is the consequence of the Frautschi picture of the Bootstrap, where the minimum number of final state hadrons are 2. This problem is solved by introducing an extra contribution to the cross section described by a Breit-Wigner cross section formula $\sigma^{BR}(E)$:
\begin{equation}
\sigma^{BR}(E)=\frac{2 s_C+1}{(2s_A+1)(2s_B+1)}\frac{4\pi}{p_i^2}\frac{E^2\Gamma_{C\rightarrow AB}\Gamma_{tot}}{(E^2-m_C^2)^2+E^2\Gamma_{tot}^2}
\label{eq:br}
\end{equation}
where $A,B$ are the colliding particles, $C$ is the created resonance, which could decay into some specific final state described by it's total decay width $\Gamma_{tot}$. The $s_{A,B,C}$ factors are the spins of the corresponding particles, and $p_i$ is the center of mass momentum of the  initial state. The relevant decay widths can be gathered from the Particle Data Book (PDG) \cite{PDG}. With this in mind the probability of a specific channel corresponding to the inelastic cross section is expressed in Eq.\ref{eq:12}.
\begin{equation}
\label{eq:12}
\frac{\sigma_i(E)}{\sigma_{inelastic}(E)} = R_i(E) + \frac{\sum_j \sigma_j^{BR}(E)}{\sigma_{inelastic}(E)}
\end{equation}
where $R_i(E)$ is the cross section ratio from our statistical model, where all the possible fireball decay probabilities are summed over (Eq.\ref{eq:13}).
\begin{equation}
\label{eq:13}
R_i(E) = \sum_{k=1}^N W_{k,i_1,..i_k}^{n_{i_1}..n_{i_k}}(E)
\end{equation}
where the sum of all the possible $k=1..N$ fireball schemes for a specific final state are taken. An example is the final state $3\pi^+ 3\pi^-$ where one has to sum the 1-,2-, and 3 fireball contributions as well. If one wants to calculate inelastic cross sections the full normalization sum has to be calculated and possibly an uncertainty analysis has to be done, which is the main focus of the next section.

\section{Normalization and uncertainty estimation}
\label{sec:2}
The normalized probability in Eq.\ref{eq:4} should give back the $\sigma_I/\sigma_{inelastic}$ cross section ratio by definition. If we have a specific channel $"i"$ which only could come from a one fireball decay scheme, then only the one fireball normalization sum should be calculated. The following four processes were calculated to test the method:
\begin{itemize}
\item $p \bar{p} \rightarrow \pi^+ \pi^-$
\item $p \bar{p} \rightarrow n \bar{n}$
\item $p \bar{p} \rightarrow \Delta^{++} \Delta^{--}$
\item $p \bar{p} \rightarrow K^+ K^-$
\end{itemize}
As the measured cross sections and the model also could contain errors it is possible to estimate an error distribution of the model from the distribution of the relative errors. To this purpose, the relative error is calculated as it is shown in Eq.\ref{eq:14}. 
\begin{equation}
\label{eq:14}
k = \frac{r_0-W_{1}}{r_0},
\end{equation}
where $r_0=\sigma_i/\sigma_{inel}$ and $W_{1}$ is the one fireball model probability. The model calculations does not contain $A+B \rightarrow C \rightarrow D+E$ reactions, due to the lack of knowledge of the branching ratios of heavy mesons to proton and antiproton. Using the Gaussian error propagation formula, considering that $\Delta \sigma_i \neq 0$, $\Delta \sigma_{inel} \neq 0$, and $\Delta W_{1} \neq 0$ the absolute error of the measured relative error $\Delta k$ is expressed in Eq.\ref{eq:15}.
\begin{equation}
\label{eq:15}
\Delta k = \frac{W_{1}}{r_0} \Big(\frac{\Delta r_0}{r_0} + \frac{\Delta W_{1}}{W_{1}} \Big)
\end{equation}
The relative error of the model could be expressed from Eq.\ref{eq:15} as follows (Eq.\ref{eq:16}):
\begin{equation}
\label{eq:16}
\frac{\Delta W_{1}}{W_{1}}  = r_0 \frac{\Delta k}{W_{1}} - \frac{\Delta r_0}{r_0}
\end{equation}
If the $\Delta k$ uncertainty is approximated by the standard deviation of the relative error distribution, and assuming an energy-dependent model error, the relative model error distribution could be expressed from Eq.\ref{eq:16} and its histogram can be seen in Fig.\ref{fig:2}:

\begin{figure}
\resizebox{0.5\textwidth}{!}{%
  \includegraphics{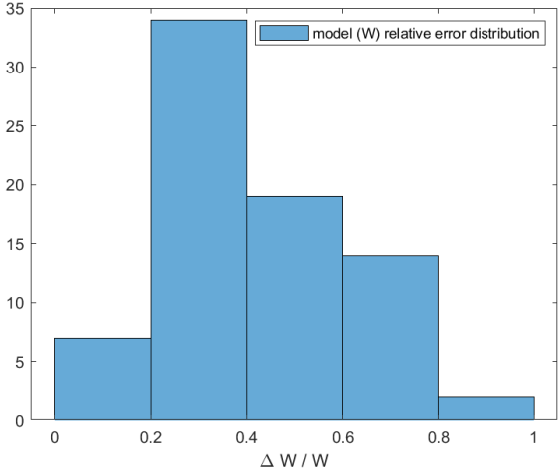}
}
\caption{Estimated relative error distribution of the model.}
\label{fig:2}
\end{figure}

\begin{figure}
\resizebox{0.5\textwidth}{!}{%
  \includegraphics{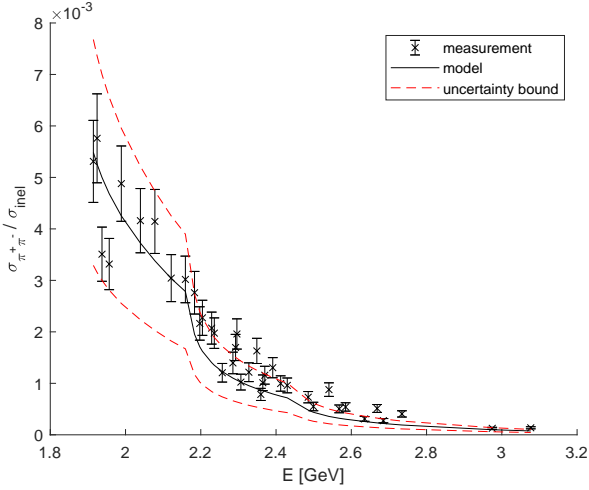}
}
\caption{The process $p \bar{p} \rightarrow \pi^+ \pi^-$ ratio to the inclusive cross section $\sigma_{p \bar{p}}^{inel}$ with the full normalization for one fireball. Data taken from \cite{exp1}\cite{exp2}.}
\label{fig:3}       
\end{figure}

The estimated relative error of the model is calculated from this distribution by taking its mean value, which is approximately $\Delta W_{1}/W_{1} \approx 0.4$. This value will be the approximated relative error of the model, which will be used in the following sections, where the inelastic cross sections are calculated. The calculated model probability with the measured ratios can be seen in Fig.\ref{fig:3}, where the uncertainty bound is also calculated from the previously estimated model error. It is again worth to mention, that in this calculations the possible $p\overline{p} \rightarrow X \rightarrow \pi^+ \pi^-$ processes are neglected, where $X$ is a heavy meson above the two proton threshold. This is mainly due to the lack of knowledge of the branching fractions at the PDG for heavy meson decays to proton and antiproton.

\section{Inclusive cross sections}
\label{sec:3}
In heavy-ion simulations, if the interesting quantities are particle multiplicities, then the desirable cross sections are the inclusive ones e.g. if one interested in the charmonium spectra which is measured by their dilepton pair decays an interesting background process could be $X \rightarrow D \bar{D}$, where each D meson could decay into an electron(positron) and a Kaon, giving an overall dilepton pair at the end. In this example, the ($X \rightarrow D \bar{D} + $ anything) process should be calculated as every possible channel is giving an extra contribution to the full background. The naive way to calculate inclusive cross sections to sum over all the possible reactions, which respects the quantum number conservation laws, however, this method is not very efficient and there is a much easier way to do this.

Let us take the probability ratio of a channel with two different normalization, which only means the normalization sum contains different channels. The inelastic sum contains all the possible final states allowed by quantum number conservation, while the inclusive sum contains every possibility, but with one specific particle fixed in every possible final state. To make things easy the reference channel should be strictly $1, 2, 3,...N$ fireball channel without mixing so that the normalization is clear. Let us take a reference channel $I$, which is coming from strictly one fireball channel and calculate the ratio like in Eq.\ref{eq:17}:
\begin{equation}
\label{eq:17}
\frac{\sigma_I / \sigma_{inelastic}}{\sigma_I / \sigma_{inclusive}} = \frac{\sigma_{inclusive}}{\sigma_{inelastic}}=\frac{\sum_{j \in inclusive}T_j(E)}{\sum_{j \in inelastic}T_j(E)}
\end{equation}
where in the ratio the common factors e.g. $P_1^{fb}$ are dropped out and only the normalization terms in the denominator remained. The Breit-Wigner factors are also considered in the phase-space integrals. In this way the inclusive cross section can be expressed as it is shown in Eq.\ref{eq:18}:
\begin{equation}
\label{eq:18}
\sigma_{inclusive} = \sigma_{inelastic} \frac{ \sum_{j \in inclusive} T_j(E) }{\sum_{j \in inelastic}T_j(E)}
\end{equation}
A problem with using the one-fireball ratio is the smallness of $\sum T_i$ at higher energies, which means that the probabilities at higher energies for all of the one fireball possibilities are negligible or practically zero. A solution to this problem is to use the two-fireball ratio due to their non-negligible normalization sums even at the energies up to 15 GeV. The inclusive cross section from strictly two-fireball channels can be expressed in the same way with the corresponding normalization sums as in Eq.\ref{eq:19}:
\begin{eqnarray}
\label{eq:19}
& {\displaystyle \sigma_{inclusive} = \sigma_{inelastic} } \times \\
& \displaystyle{ \times \Bigg[ \sum_{ij \in S_{inclusive}} \frac{1}{N_{ij}!} \int{dx H_i(x)H_j(E-x)}\Bigg] } / \nonumber \\ 
& \displaystyle{  / \Bigg[ \sum_{kl \in S_{inelastic}} \frac{1}{N_{kl}!} \int{dx H_k(x)H_l(E-x)}\Bigg]} \nonumber
\end{eqnarray}
where $\sum_{ij \in S_{inclusive}}$ means every possible two-fireball final state from the inclusive set, which respects the conservation laws of the initial state, and for simplicity, we defined $H_i(x)$ as:
\begin{equation}
\label{eq:20}
H_i(x) = \frac{T_i(x)}{\sum_{j}T_j(x)}
\end{equation}

Typically inclusive cross sections like $p \bar{p} \rightarrow I+X$ are needed where $I$ is a fixed particle. The inclusive sum then includes all of the possible 2 fireball final states, which contains at least once particle $I$ in its final state, however, it is also possible to calculate inclusive reactions like $p \bar{p} \rightarrow I+J+X$, where $I$ and $J$ are two fixed particles and $X$ is everything else. The generalization is straightforward and the only task is to separate the possible 2 fireball reactions where we have a specific number of fixed particles. For one fireball scheme, the calculation is simple and it is possible to include every possible two and three-body final states into the sum with little effort, however, if one wants to calculate the ratio from more than one fireball decays the possible number of channels are too huge and the integration would take many hours on a standard laptop. Due to the numerical complexity, except for one case, only the one fireball calculations are shown in this paper, however, it can be shown that if one wants to lower the uncertainty of the model, then taking the two fireball normalization ratio is a good option to do so. To see this let us assume that each $H_i(x)$ is a uniformly distributed random variable with some mean $\mu>0$ and a relative width $0.5\mu$. Taking the two-fireball normalization ratio the following distribution has to be calculated (Eq.\ref{eq:21}):
\begin{equation}
\label{eq:21}
f(E) = \frac{{\displaystyle\int dx \sum_{i,j=1}^{M} g_i(x)g_j(E-x)}}{{\displaystyle\int dx \sum_{i,j=1}^N g_i(x)g_j(E-x)}},
\end{equation}
where $g$ is a uniformly distributed random variable, and $M<N$ because $M$ corresponds to the inclusive sum, and $N$ corresponds to the inelastic sum. For simplicity let us assume that each $g_i$ is energy independent so that $f(E)=f$ will also be energy independent. To further simplify the problem the ratio in Eq.\ref{eq:22} is calculated:
\begin{equation}
\label{eq:22}
f = \frac{ \sum_{i=1}^{M} g_i^2}{ \sum_{i=1}^N g_i^2},
\end{equation}
where $g_i$ is now constant, so the integral can be factored out, and instead of taking every possible $i,j$ combinations, we simply take the square of the generated $g_i$ random variables. This will not change the main conclusion, so for the first estimation, it is sufficient to see the relative error distribution of the calculated ratio in Eq.\ref{eq:22}. The relative error distribution is calculated in the following steps:
\begin{enumerate}
\item Generate $N$ number of uniformly distributed $U[0,1] \rightarrow g_i^0$ samples. These will be the noiseless data.
\item Add a random noise to each of the $g_i^0$'s so that $g_i=g_i^0 + U[g_i^0-0.5g_i^0,g_i^0+0.5g_i^0]$ will be the data with a relative error of 0.5.
\item Take $M$ number of samples from the previously generated samples. These will correspond to the inclusive sum.
\item Calculate the ratios with the noiseless samples $f_0$, and with the noisy ones $f$.
\item Calculate the relative error $r=|f_0-f|/f_0$
\end{enumerate}

\begin{figure}
\resizebox{0.5\textwidth}{!}{%
  \includegraphics{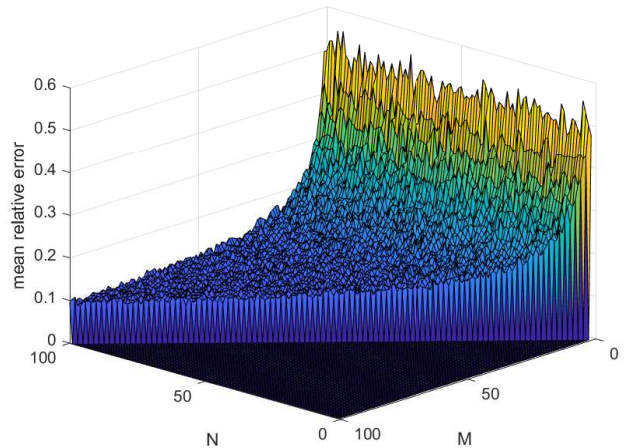}
}
\caption{The relative error dependence of the two fireball normalization ratios on the number of inelastic-, and inclusive channels (N,M).}
\label{fig:4}       
\end{figure}

Do the previous steps many times so at the end a relative error distribution from the calculated $r$'s is obtained. As has been done previously, the estimated relative error will be the mean of the obtained relative error distribution. The mean relative error dependence on the number of inelastic channels (N) $i=1..100$, and the number of inclusive channels (M) $j=i..100$ can be seen in Fig.\ref{fig:4}. The number of inelastic channels has to be larger than the number of inclusive ones e.g. $N>M$. The relative error at low multiplicity is around 0.5, which means the accumulated relative error is the same as the individual relative errors. If the multiplicity in the numerator and in the denominator increasing the accumulated relative error will be smaller. The main conclusion from this simple simulation is that if one wants to make the estimated relative error smaller it is sufficient to calculate e.g. the two fireball normalization sums instead of the one fireball sums, however, its numerical complexity is much greater. Also, the two fireball cases will cover a different energy range due to the different thresholds and different processes considered.

\section{Results}
\label{sec:4}
\subsection{Non-charmed hadrons}
\label{sec:41}
\begin{figure}
\resizebox{0.5\textwidth}{!}{%
  \includegraphics{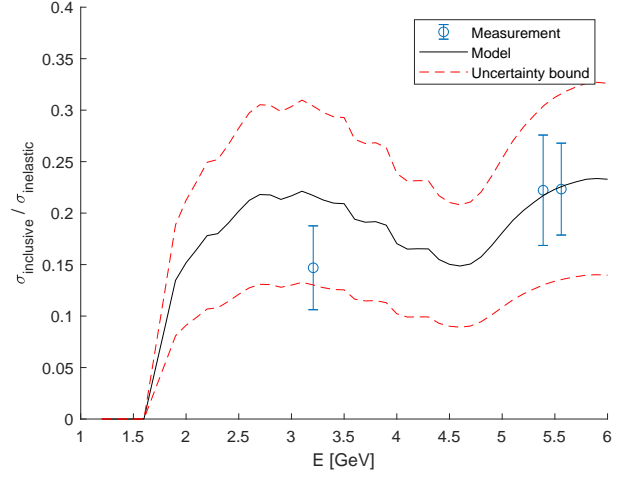}
}
\caption{The process $p \pi^- \rightarrow \rho^0 X$ ratio to the inclusive cross section $\sigma_{p \pi^-}^{inel}$. Data taken from \cite{exp1}\cite{exp2}.}
\label{fig:5}       
\end{figure}

To validate the theory six channels were calculated with the formalism showed in Sect.~\ref{sec:2}. In three process the one fireball probability ratios were calculated and compared to the measured cross section ratios, while in one case the two fireball ratio is calculated. The errors of the measured- and the calculated cross section ratios are also shown. The first reaction is the $p \pi^- \rightarrow \rho^0 X$, where $X$ is any $1,2,...N$ number of particles, which allowed by the quantum number conservation of the initial $p\pi^-$ collision, and is allowed by the specific number of fireballs we used in the calculations. To calculate the ratio, first, the inelastic sum has to be done. The second step is to find all the resonances which could decay into $\rho^0$ e.g. $N_{1700} \rightarrow \rho N$ and select only those channels into the sum (with the direct $\rho^0 X$ channels as well). These are most of the $\Delta$-s and nucleon resonances and are taken from the PDG, where only particles that have at least 3 stars are included. The results can be seen in Fig.\ref{fig:5}.

\begin{figure}
\resizebox{0.5\textwidth}{!}{%
  \includegraphics{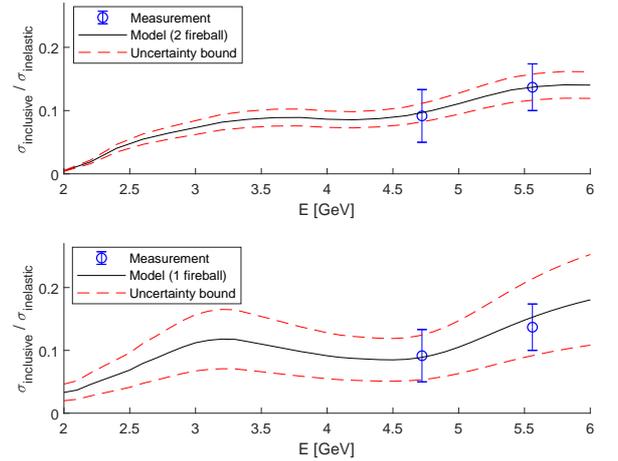}
}
\caption{The process $p \pi^- \rightarrow K^0 X$ ratio to the inclusive cross section $\sigma_{p \pi^-}^{inel}$ from two- and from one fireball processes. Note the smaller uncertainty bound in the two fireball case. Data taken from \cite{exp1}\cite{exp2}.}
\label{fig:6}       
\end{figure}

In the second example, the process $p \pi^- \rightarrow K^0 X$ is calculated with one- and also with two fireball ratios in order to compare the results. In the inclusive sum, the particles which could decay into $K^0$-s have to be included. These are some of the $N,\Lambda$, $\Sigma$, and $K^*$ resonances. The inelastic sum has to include all processes which have one baryon number, zero strangeness, zero charmness, and zero charge. The ratios from the two- and one fireball processes are shown in Fig.\ref{fig:6}, where in both cases a really good agreement with the measured data is achieved. The error in the two fireball case is calculated from the simulation described in Sec.\ref{sec:3}, using the number of inelastic, and the number of inclusive channels.

\begin{figure}
\resizebox{0.5\textwidth}{!}{%
  \includegraphics{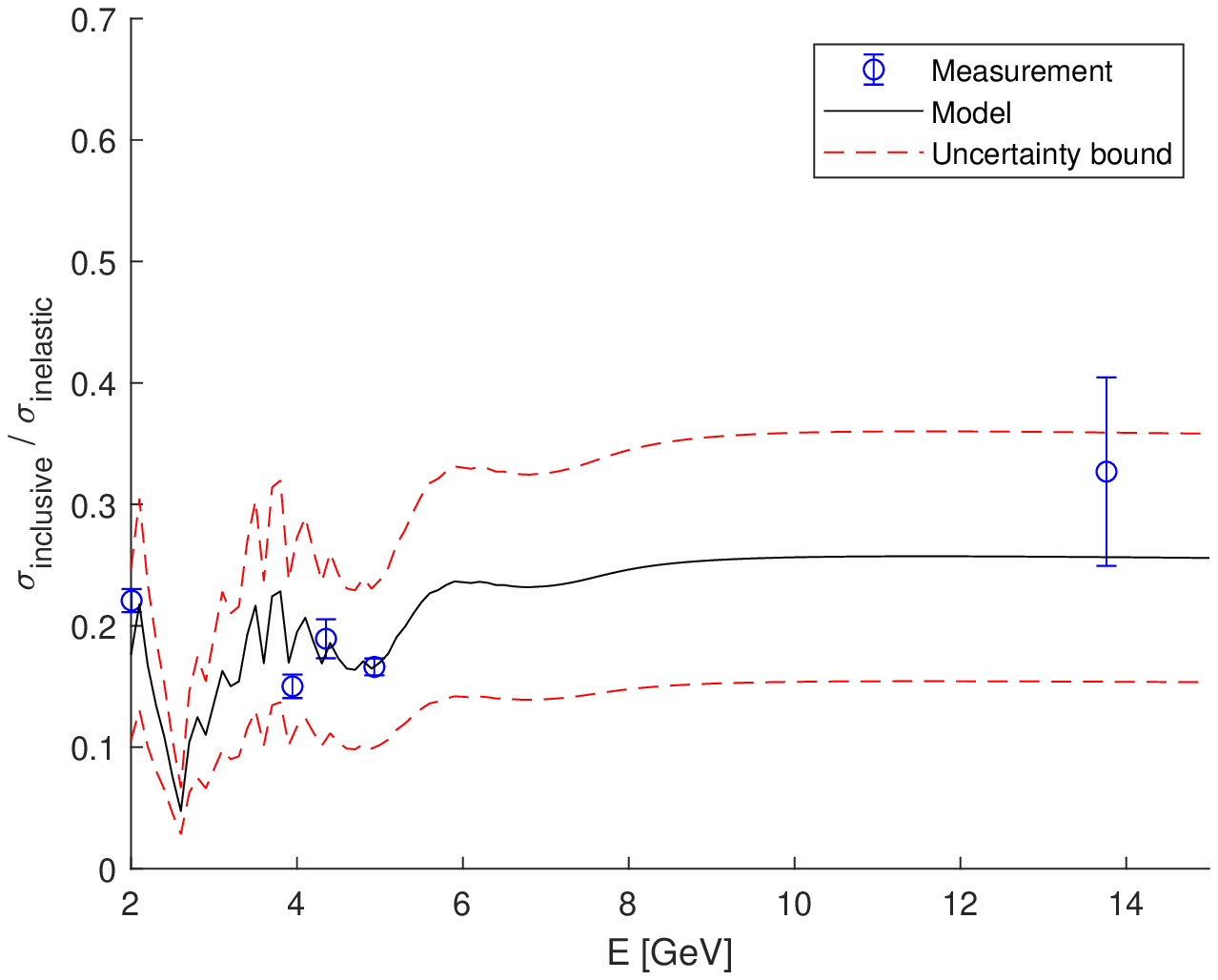}
}
\caption{The process $p \bar{p} \rightarrow \rho^0 X$ ratio to the inclusive cross section $\sigma_{p \bar{p}}^{inel}$. Data taken from \cite{exp1}\cite{exp2}.}
\label{fig:7}       
\end{figure}

The third example is the channel $p \bar{p} \rightarrow \rho^0 X$, where the inelastic sum will be different than in the previous two reactions because in $p \bar{p}$ collisions all the important conserved quantum numbers are zero. The results are shown in Fig.\ref{fig:7}, where again a really good agreement is achieved even at higher energies (15 GeV).

\begin{figure}
\resizebox{0.5\textwidth}{!}{%
  \includegraphics{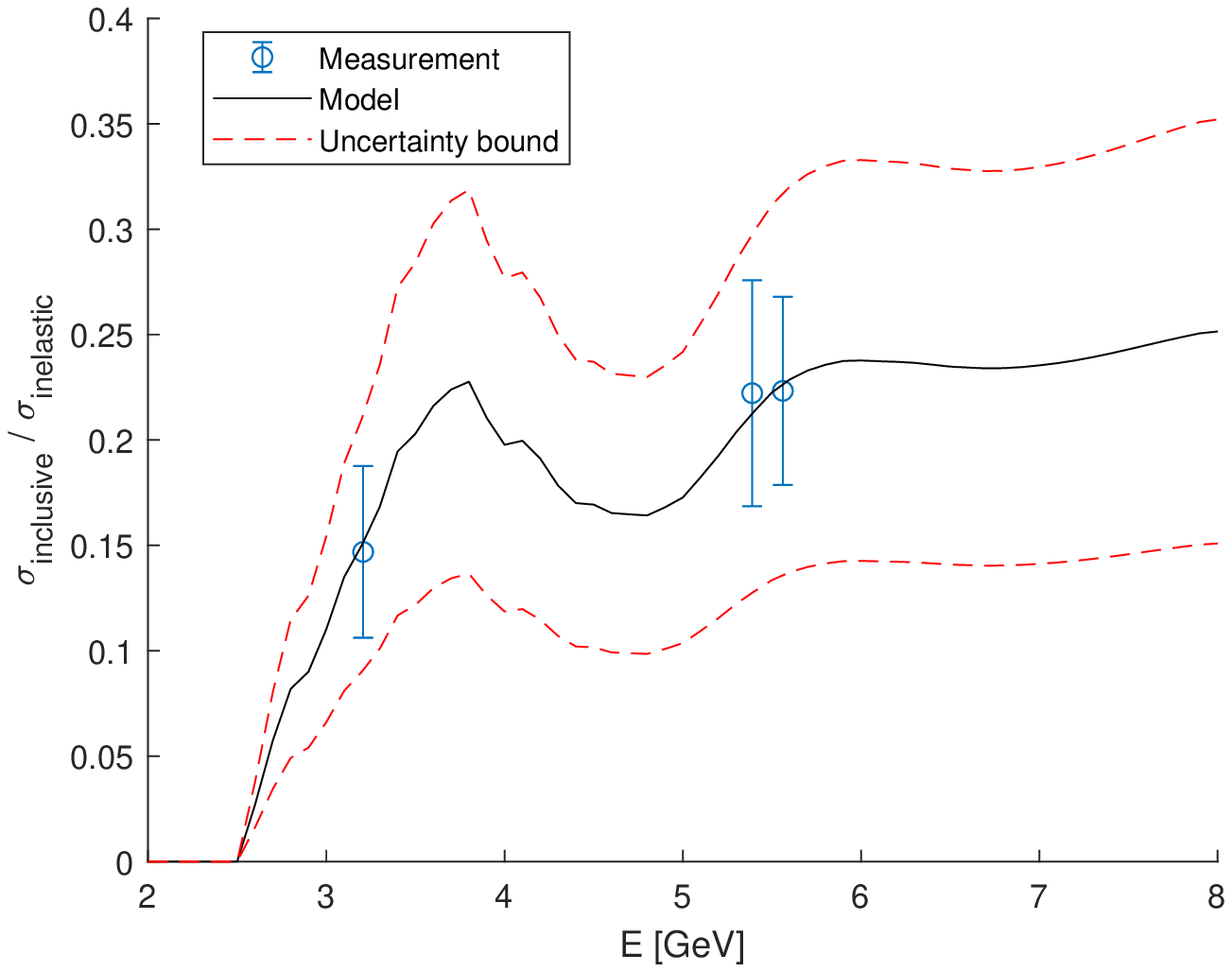}
}
\caption{The process $p p \rightarrow \rho^0 X$ ratio to the inclusive cross section $\sigma_{p p}^{inel}$. Data taken from \cite{exp1}\cite{exp2}.}
\label{fig:8}       
\end{figure}

The fourth calculated process is the $p p \rightarrow \rho^0 X$ inclusive channel, where the inelastic sum includes all the possible final states with quantum numbers (baryon number, charge, strangeness, charmeness)=(2,2,0,0). The inclusive sum contains all the processes where at least one $\rho^0$ meson is present and also all the relevant resonances which could decay to $\rho^0$. The results are shown in Fig.\ref{fig:8}.

\begin{figure}
\resizebox{0.5\textwidth}{!}{%
  \includegraphics{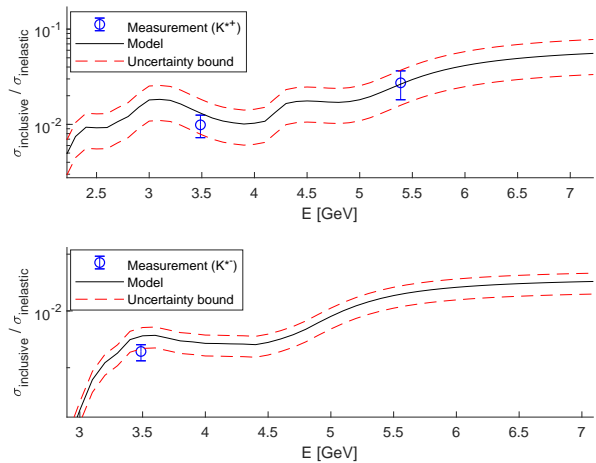}
}
\caption{The processes $p \pi^- \rightarrow K^{*}(892)^+ X$ and $p \pi^- \rightarrow K^{*}(892)^- X$ ratio to the inclusive cross section $\sigma_{p \pi^-}^{inel}$. The upper half of the figure shows the $K^{*}(892)^+$ production, while the the lower half shows the $K^{*}(892)^-$ case. Data taken from \cite{exp1}\cite{exp2}.}
\label{fig:9}       
\end{figure}

The final non-charmed examples are the strange vector meson $K^{*}(892)^+$  and $K^{*}(892)^-$ production in $\pi^- p$ collisions. The strange vector meson production is an important tool to study the dense hadronic matter in heavy ion collisions \cite{kaon1}\cite{kaon2}\cite{kaon3} due to their spectral function dependence on the nuclear density and temperature. We expect the model to give back the low energy suppression of the $K^{*}(892)^-$ state to the $K^{*}(892)^+$ meson. The results can be seen in Fig.\ref{fig:9}, where the upper side shows the $K^{*}(892)^+$ ratio to the $\pi^- p$ inelastic cross section and the lower side shows the $K^{*}(892)^-$ one. Also some measured values are shown from \cite{exp1}\cite{exp2}. It can be deduced that the main feature ($K^{*}(892)^-$ suppression) can be described quite well with the model. It should be worth noted however that for the strange quarks, we used a constant value for the $P_s$ quark creational probability, fitted from $p \bar{p} \rightarrow K^+ K^-$ process at the energy range $E \approx [1.8,2.6]$ GeV, which is a simplification, as the strange quark suppression should be energy dependent, so does $P_s$. Nevertheless, even with this assumption the model gave back the main qualities of the cross sections.

In every aforementioned process, a really good match is achieved with only one fireball process considered. Using two fireball ratios the match is even better, which corresponds to the smaller model error estimated in the previous section.


\subsection{Charmonium production}
\label{sec:42}
One of the models aim is to give estimates to charmonium creation cross sections below 10 GeV, which will be very useful inputs for upcoming transport simulations, regarding charmonium mass shifts in heavy ion collisions. To make reliable estimates the model should describe well the available low energy cross sections, which are very rare at the moment. A good collection of data are used for fitting in \cite{charm1}. In our model, the necessary parameter, which has to be fitted from experiments is the charm quark creational probability ($P_c$), which is used to calculate the quark combinatorial factors. Due to the much larger mass of the charm quark than the u,d,s quarks, this value are expected to be much smaller then the others, thus to create more than one charm-anticharm pair is negligible in the calculations. Furthermore we expect it to increase with energy, rather than to be a constant value. In contrast with our previous attempt in \cite{Balassa:2017pgf}, where we tried to fix a constant $P_c$ value at one point using a fit from \cite{charm1} near threshold, with the calculation of one exclusive channel, now we have the method to fit for measured inclusive data at higher energies. This was one of the main reasons we extended the model to describe inclusive cross sections. 

For the quark creation probability, we expect a linear rise with energy with the simplest assumption possible $P_c=aE$, where only one parameter, the slope $a$ has to be determined. This parameter enters into the quark number probability mass function described by a multinomial distribution shown in Eq.\ref{eq:23}
\begin{equation}
\label{eq:23}
F(N,n_i) = \frac{N(E)!}{\prod_{i=u,d,s,c}n_i!} \prod_{i=u,d,s,c} P_i^{n_i},
\end{equation}
where 
\begin{equation}
\label{eq:24}
N(E)=\frac{1+\sqrt{1+E^2/T_0^2}}{2},
\end{equation}
is the total number of quark-antiquark pairs \cite{chiral}, with hadronization temperature $T_0$, and $P_i$ is the quark creational probability for the $i=(u,d,s,c)$ type quarks. The expected number of quark-antiquark pairs of type $i$, will be $\overline{n}_i=P_iN$, which corresponds to the maximum of the probability mass function. Note that the bottom quark is still missing, and we intend to fit the corresponding $P_b$ value as well in the future. The expected number of different quarks and antiquarks then build up the hadrons in all possible ways described by simple combinatorics, multiplied by the corresponding probability from the probability mass function, and at the end are normalized with all the possible final states, giving a final probability to create one specific hadronic final state. As the charm quark creation is expected to be highly  suppressed only $\overline{n}_c=1$ is interesting at the moment, which will corresponds to a constrained maximum value of $F(N,n_i)$ in contrast to the non-constrained maximum, where $\overline{n}_c=0$. The suppression ratio is then proportional to the charm-anticharm creational probability, so it can be easily fitted using the suppression to the non-charmed channels. To see this, let us assume that the one charm-anticharm quark pair is created in expense to one strange-antistrange quark pair, so $\overline{n}_c'=1$, and $\overline{n}_s'=\overline{n}_s-1$, while $\overline{n}_u'=\overline{n}_u$ and $\overline{n}_d'=\overline{n}_d$. The suppression ratio calculated from the probability mass function is then:
\begin{equation}
\label{eq:25}
\gamma_c = \frac{F(N,\overline{n}_u',\overline{n}_d',\overline{n}_s',\overline{n}_c')}{F(N,\overline{n}_u,\overline{n}_d,\overline{n}_s,\overline{n}_c)} = \frac{\overline{n}_s! P_c}{(\overline{n}_s-1)! P_s}=P_c N
\end{equation}
The derived formula for $\gamma_c$ expresses the fact, that even with a constant $P_c$ value, the suppression will be energy dependent if $\gamma_c<1$, as the total number of quark-antiquark pairs $N$ are also energy dependent. After $\gamma_c$ reaches 1 the suppression from the probability mass function is gone, so no need to constrain the distribution anymore. The expected number of charm-anticharm quarks then will be $\overline{n}_c=P_c N$, so the transition between the suppressed and non-suppressed probability is continous. The vanishing suppression ratio here only means, that there will be no more suppression from the constrained probability mass function, however if $P_c$ is still smaller than the other quark creational probabilities (which will be the case up to a few 100 GeV), then the charm production is still suppressed compared to the up, down and strange quarks. 

As we assumed a linear relationship between $P_c$ and the energy, the $\gamma_c$ suppression ratio can be expressed as:
\begin{equation}
\label{eq:26}
\gamma_c = a \frac{E+E \sqrt{1+E^2/T_0^2}}{2},
\end{equation}
where $a$ is the slope parameter, which has to be determined from experiments. Using experimental charmonium production cross section data in $pN$ collisions, this parameter is determined to be $a=0.0006 $ GeV$^{-1}$ (see below), so the charm-anticharm quark creational probability is given by $P_c=0.0006 \cdot E$. With the inclusion of another type of quark the previous $P_u, P_d, P_s$ fits were also have to be modified. In the simplest case, we can assume that $P_i'=P_i-P_c/3$ for $i=(u,d,s)$, so that $P_u'+P_d'+P_s'+P_c=1$ holds, however due to the small value of $P_c$ in the energy range below a $100$ GeV, we get $P_u' \approx P_u$, $P_d' \approx P_d$, $P_s' \approx P_s$, so the previously used values for the $u,d,s$ quarks can still be used in this energy range.

\begin{figure}
\resizebox{0.5\textwidth}{!}{%
  \includegraphics{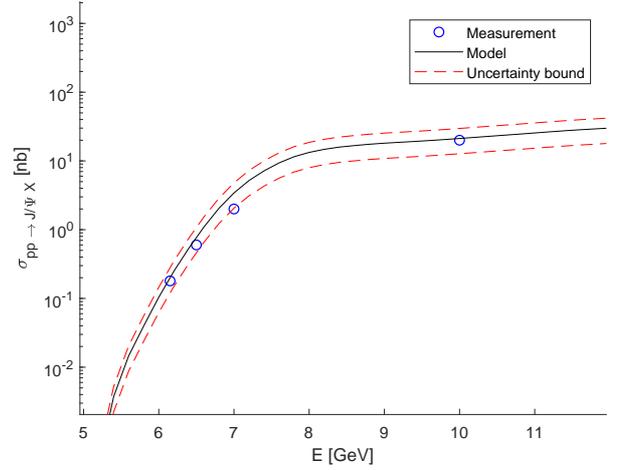}
}
\caption{Inclusive $J/\Psi$ production cross section estimates in proton-proton collisions. Only the rightmost measured point near $10$ GeV is used to fit the slope in $P_c=aE$. Data taken from \cite{charm1}.}
\label{fig:10}       
\end{figure}

\begin{figure}
\resizebox{0.5\textwidth}{!}{%
  \includegraphics{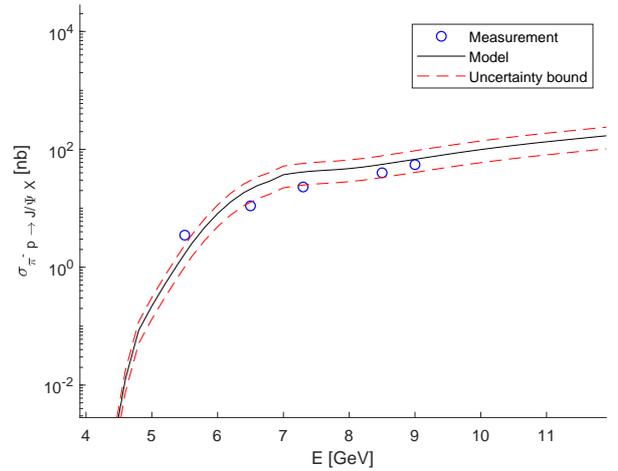}
}
\caption{Inclusive $J/\Psi$ production cross section estimates in $\pi^- p$ collisions. Data taken from \cite{charm1}.}
\label{fig:11}       
\end{figure}

To calculate the cross sections the two fireball probabilities are used, as it gives smaller uncertainties, with the price of taking much longer time to calculate the normalization factors. We intend to describe the available low energy $pN$ and $\pi N$ data with the model for which, we selected some experimental points below $10$ GeV from the collection in \cite{charm1}. At this stage the uncertainties in the experiments are not really important as the main goal is to describe the 2-3 times larger cross sections of the $\pi N$ collisions in contrast to the $p N$ collisions. To this end, and to check the model predictability, only one measurement point is used to the fit from the $pN$ data at $E\approx 10$ GeV, which is enough, as we only have one free parameter in $P_c$, namely $a$. The tuning was done by hand to approximately give the correct value at the one point used for the fit. For the cross sections the decays from $\Psi(3686)$, $\chi_{c1}$ and $\chi_{c2}$ mesons are also included, however as there are no bottom quarks in the current model the decays from hadrons containing $b$ quarks are not included. After the tuning of the slope parameter, the full cross sections were calculated for the $pp$ and for the $\pi^- p$ case. The results can be seen in Fig.\ref{fig:10} and in Fig.\ref{fig:11}. The calculations show a good match with the data for both cases even with the simple form of the $P_c$ and with the restricted one point fit, thus it can be concluded that the main feature of the $\pi N$ enhancement is described well in the model.
To test the model capabilities further the proton-antiproton to proton-proton inclusive $J/\Psi$ cross section ratio is calculated at $E=24.3$ GeV CM energy, which is measured in\cite{charm2} with a good accuracy. The results with the uncertainties are shown in Table.\ref{Table:ratio}, where a remarkably good agreement with the measured value is achieved.

The method thus proved to be able to give satisfactory results to low/medium energy charmonium cross sections, however putting more particles (e.g. B mesons) into the model could change the higher energy parts of the cross sections, meaning the $P_c$ quark creational probabilities might have to be further fine tuned. Also for $P_c$ a very simple linear assumption was made, and it could be interesting to give a theoretical background to its nature. This also applies to the strange quark creational probability $P_s$, where we simply assumed a constant value throught the whole energy range we are interested in. 

\begin{table}[]
\centering
\begin{tabular}{|l|l|l|}
\hline
     & Measurement        & Model \\ \hline
ratio & $0.76 \pm 0.14 \pm 0.06$ &   $0.73 \pm 0.15 $    \\ \hline
\end{tabular}
\caption{Inclusive $J/\Psi$ cross section ratio for ($pp \rightarrow J/\Psi X$) / ($\bar{p}p \rightarrow J/\Psi X$) collisions at $\sqrt{s}=24.3$ GeV. Data is taken from \cite{charm2}.}
\label{Table:ratio}
\end{table}

\section{Conclusion}
\label{sec:5}
In this paper, a statistical model is introduced based on the statistical Bootstrap approach, which is able to give reasonable good estimates to hadronic cross section ratios up to a few GeV energy range. The full normalization sum is calculated for one and two fireball decay schemes. The cross section ratios for eight different processes were compared to the model calculations. From the calculated relative errors, the model uncertainty is also estimated giving an overall energy independent relative error for the one fireball decay probability. The calculated normalized probabilities are in good agreement with the measured values. The model is extended to give inclusive cross sections from the ratios of the normalization sums, which is validated through six distinct inclusive non-charmed processes, namely, $p \pi^- \rightarrow \rho^0 X$, $p \pi^- \rightarrow K^0 X$, $p \pi^- \rightarrow K^{*+} X$, $p \pi^- \rightarrow K^{*-} X$, $p p \rightarrow \rho^0 X$ and $p \bar{p} \rightarrow \rho^0 X$, where in each case a really good match with measured data is achieved. The charm quark creational probability is fitted with the help of the $ p N \rightarrow J/\Psi X$ inclusive data, allowing us to give estimates to charmed final states as well. The model is further validated through two charmed processes, namely $p p \rightarrow J/\Psi X$ and $\pi^- p \rightarrow J/\Psi X$, both giving good results compared to the experimental data. For a last application the ($pp \rightarrow J/\Psi X$) / ($\bar{p}p \rightarrow J/\Psi X$) ratio is calculated at  $\sqrt{s}=24.3$ GeV giving a really good match with the experimental data. The method will be highly useful in describing previously unknown or not well measured inclusive cross sections e.g. charmonium creational probabilities and will be used in a BUU type \cite{BUU1,BUU2,BUU3} off-shell \cite{OFFSHELL} transport code to study the propagation of these states in dense nuclear matter.
%

%
\bibliographystyle{unsrt}
 \bibliography{Main_Document}

%
%
%


\end{document}